\documentclass{moriond}

\def\be{\begin{equation}}
\def\ee{\end{equation}}
\def\bea{\begin{eqnarray}}
\def\eea{\end{eqnarray}}

\def\pth{\ensuremath{p_\mathrm{T}^H}}
\def\njets{\ensuremath{N_{\mathrm{jets}}}}
\def\mjj{\ensuremath{m_{jj}}}
\def\ptv{\ensuremath{{p_\mathrm{T}^V}}}
\def\klambda{\ensuremath{\kappa_\lambda}}
\def\ktwov{\ensuremath{\kappa_{2V}}}

\newcommand{\Photo}{\includegraphics[height=35mm]{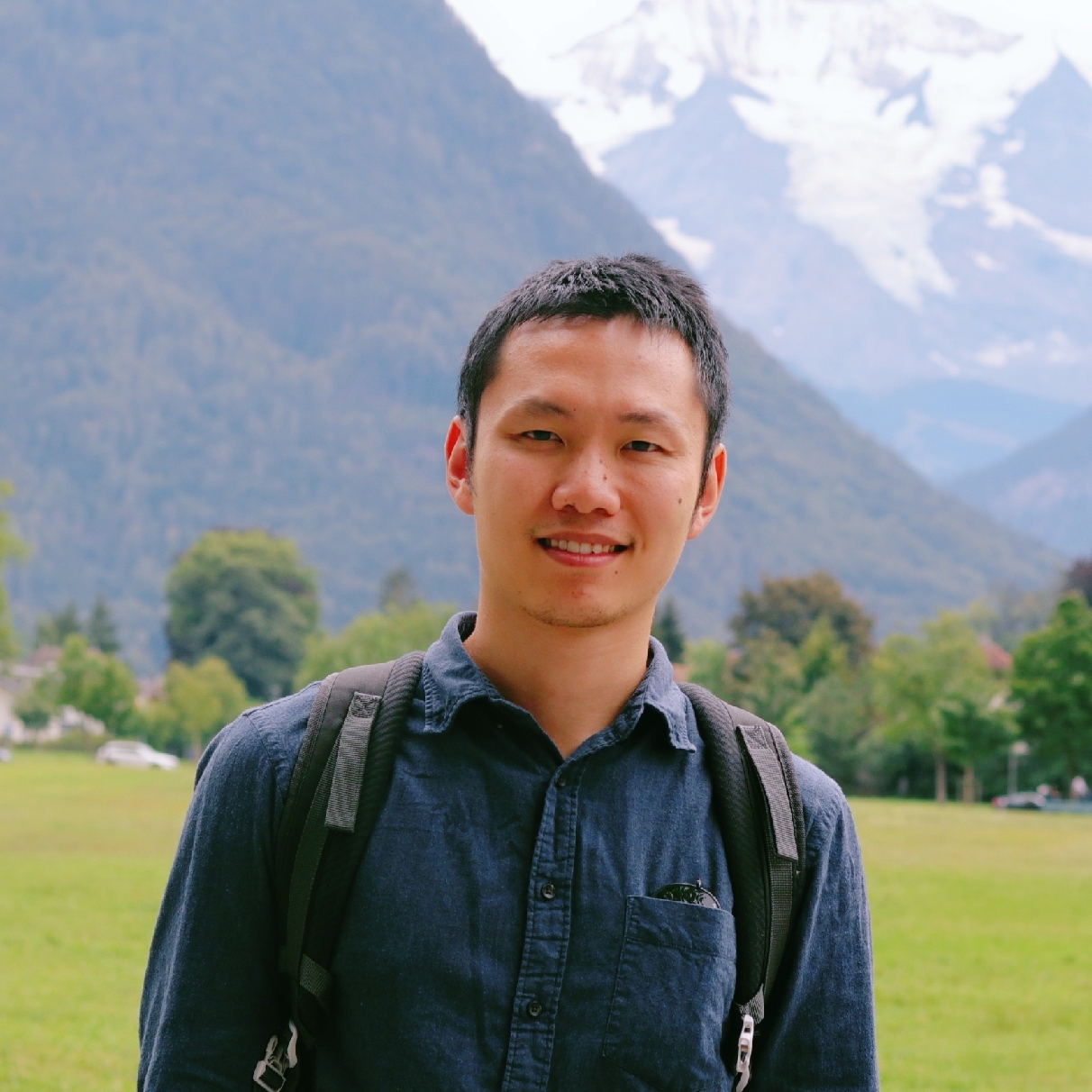}}

\begin{document}
\vspace*{4cm}

\title{Higgs cross-section (including di-Higgs) with ATLAS and CMS}

\author{Rongkun Wang \\on behalf of the ATLAS and CMS collaborations}

\address{Espl. des Particules 1,\\
1211 Meyrin, Switzerland} 

\maketitle

\footnotetext{  \copyright\ Copyright [2023] CERN for the benefit of the ATLAS and CMS Collaborations.
CC-BY-4.0 license. }

\abstracts{The latest results of Higgs cross section measurements from the ATLAS and CMS experiments were presented.
This proceeding focuses on the production cross section measurement under the simplified template cross section framework in multiple decay modes and their combination.
This proceeding also covers the differential cross section measurement and di-Higgs cross section limits and the interpretation of self-coupling constants.
Most of the results are based on the full LHC Run 2 dataset and achieve the highest sensitivity to date which marks an important milestone of Higgs physics entering the precision era.}
%
%

\section{Introduction}

The discovery of the Higgs boson~\cite{20121,201230} in 2012 has brought us new research opportunities. 
The Higgs boson production cross section multiplied by the branching ratio can be measured to probe the coupling constants of Higgs boson to other particles.
Differential cross sections as a function of kinematic variables of Higgs boson offer a more model-independent approach due to less assumption being made on the kinematic distributions.
Di-Higgs production has become more popular because it is a direct measurement of the Higgs trilinear coupling constants.
Using the full Run 2 data delivered by the Large Hadron Collider (LHC), ATLAS~\cite{ATLAS:2008xda} and CMS~\cite{CMS:2008xjf} provide the most precise Higgs cross section measurements in various Higgs decay modes and their combinations. 
This proceeding covers analyses using the full Run 2 data unless stated otherwise. 

\section{Production Cross Section Measurements}

The simplified template cross section (STXS) framework~\cite{Berger:2019aa} categorizes the cross section measurements into different kinematic regions of interest 
based on variables including Higgs boson transverse momentum (\pth), jet multiplicity (\njets), dijet invariant mass (\mjj), and also transverse momentum of the vector boson (\ptv) in \textit{VH} production mode.
The framework provides a common guideline for analyses in all decay modes for comparison and combination. Each analysis needs to merge bins when statistics is low.
The cross sections are measured simultaneously in a fit using observables which are designed for picking out events from truth bins.

\begin{figure}
\begin{minipage}{0.3\linewidth}
\centerline{\includegraphics[width=1 \linewidth]{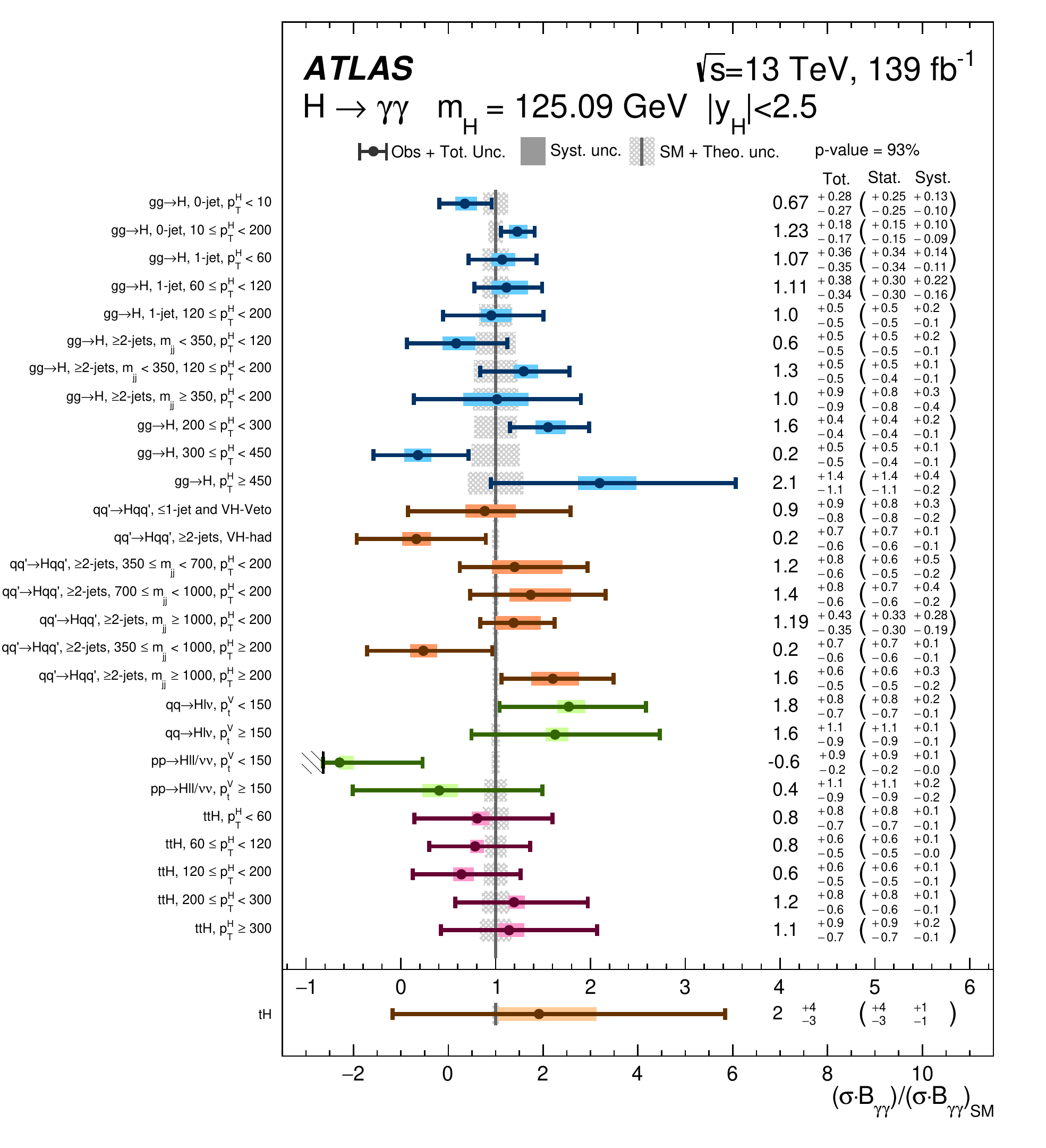}}
\end{minipage}
\hfill
\begin{minipage}{0.32 \linewidth}
\centerline{\includegraphics[width=1 \linewidth]{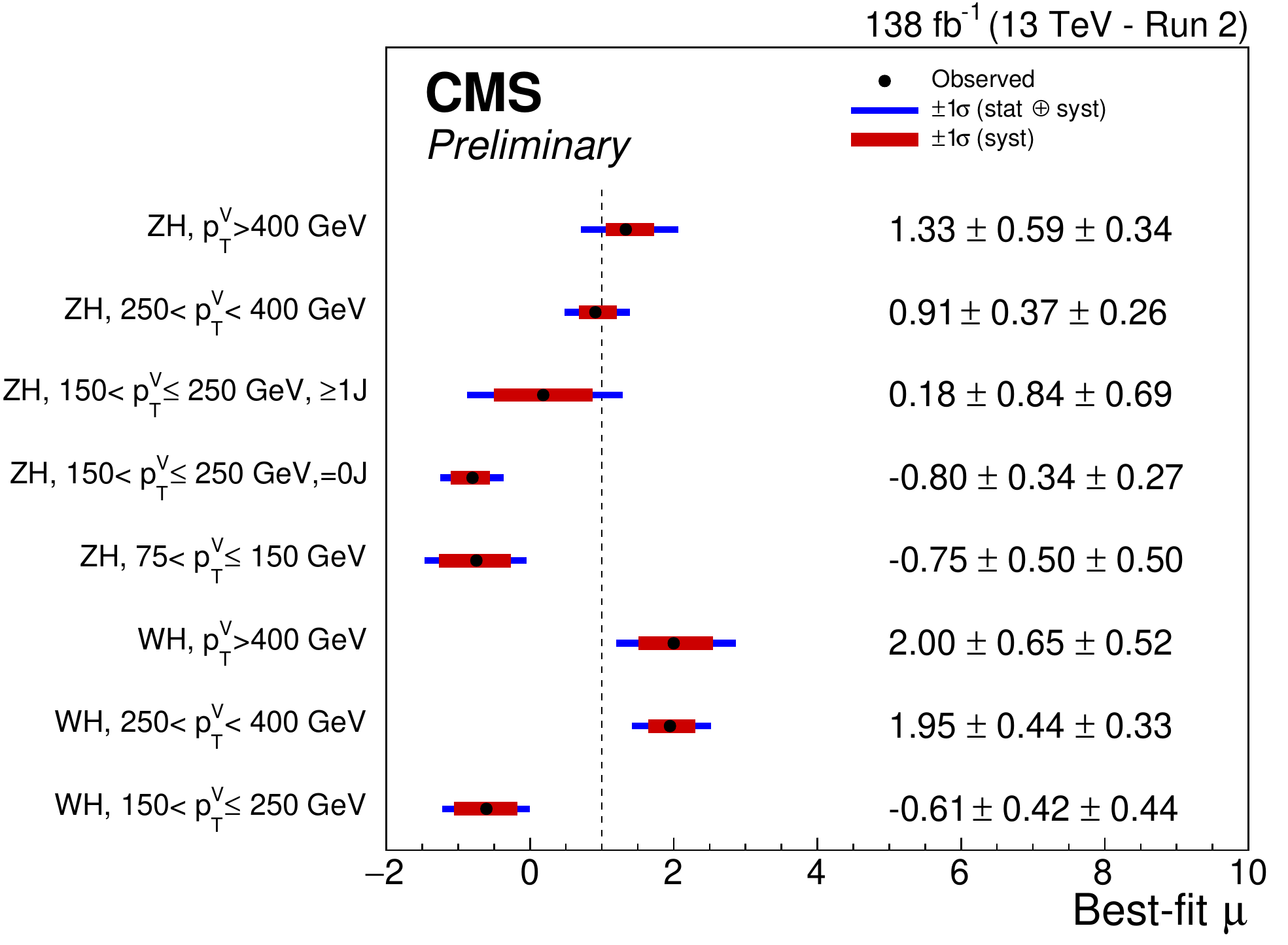}}
\end{minipage}
\hfill
\begin{minipage}{0.33 \linewidth}
\centerline{\includegraphics[width=1 \linewidth]{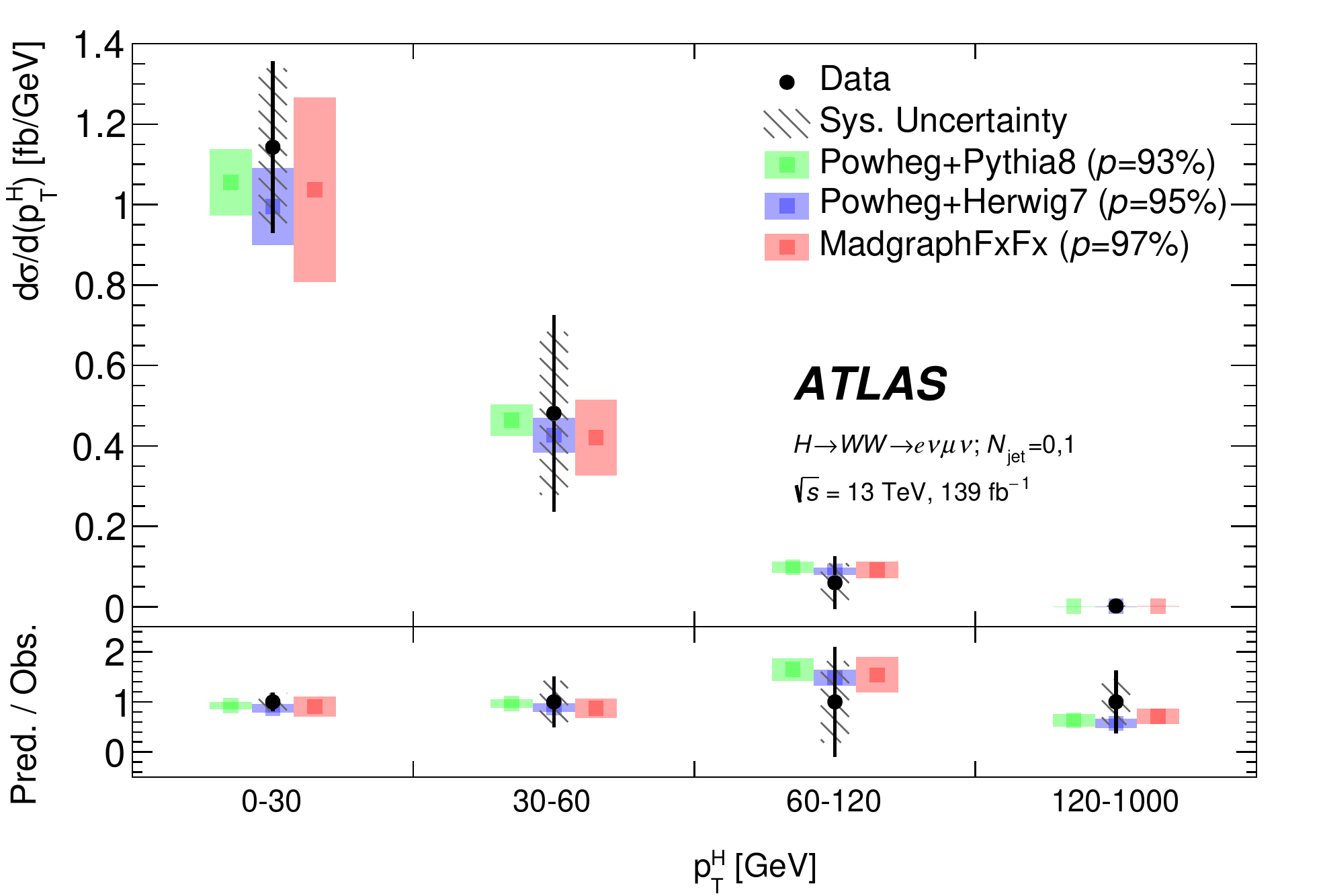}}
\end{minipage}
\hfill

\caption[]{The measured cross sections in targeted STXS bins as measured by ATLAS in the $H \rightarrow \gamma\gamma $ channel (left)~\cite{HyyA} and by CMS in the $VH \rightarrow b\bar{b} $ channel (middle)~\cite{CMS-PAS-HIG-20-001}. The differential cross sections as a function of \pth\ measured by ATLAS, in $H \rightarrow WW^*$ channel (right)~\cite{ATLAS:2023hyd}.}
\label{fig:hyyhbb}
\end{figure}

The $H \rightarrow ZZ^* \rightarrow 4\ell\ (\ell = e, \mu)$ channel~\cite{HZZA,HZZC} offers a clean signature, a high signal-to-background ratio and good statistics for the STXS measurements.
The $H \rightarrow \gamma\gamma $ channel has a fully reconstructable final state and a higher branching ratio which lead to higher statistics allowing for a much finer splitting of STXS bins, including \textit{ttH} cross sections differentially~\cite{HyyA,HyyC} in \pth as shown in Figure~\ref{fig:hyyhbb} (left). 
The total cross section of $H \rightarrow \gamma\gamma$ is also measured by ATLAS using early Run 3 data~\cite{ATLAS-CONF-2023-003}, which is consistent with the theory prediction.
The $H \rightarrow WW^*$ channel benefits from an even higher branching ratio but also suffers from higher background and non-fully-reconstructable final states. The analyses measure STXS cross sections in the ggF, VBF using fully leptonic decay mode and \textit{VH} production modes using both fully leptonic and semi-leptonic decay modes~\cite{hwwll,ATLAS-CONF-2022-067,CMS:2022uhn}.
In the $H \rightarrow b\bar{b}$ channel, the large branching ratio of about 60\% brings even more statistics.  
However, with the unique background composition in different production mode, analyses are carried out in \textit{VH}~\cite{ATLAS:2020fcp,CMS-PAS-HIG-20-001,ATLAS:2020jwz}, VBF mode~\cite{ATLAS:2020bhl,CMS-PAS-HIG-22-009} and \textit{ttH} mode~\cite{ATLAS:2021qou}. 
In \textit{VH} and \textit{ttH} modes analyses are done in \textit{resolved} and \textit{boosted} Higgs topology separately. 
In the \textit{resolved} topology the two \textit{b}-jets can be separately reconstructed.
On the contrary, in the \textit{boosted} topology the Higgs boson is boosted and the $b\bar{b}$ pair has a smaller angular separation which are reconstructed as a single large-radius jet.
For \textit{VH} analysis the boosted topology is defined as $\ptv > 250$ GeV, and for \textit{ttH} analysis it is defined as $\pth > 300$ GeV.
The \textit{boosted} analysis provides sensitivity in the phase space with \ptv(\pth) higher than $400$ GeV.
Figure~\ref{fig:hyyhbb} (middle) shows the measurement done by CMS in $VH \rightarrow b\bar{b} $ channel.
The $H \rightarrow \tau\tau$~\cite{ATLAS:2022yrq,CMS:2022kdi} analysis targets ggF, VBF, \textit{VH} and \textit{ttH} mode and offers additional sensitivity.

Both ATLAS~\cite{ATLAS:2022vkf} and CMS~\cite{CMS:2022dwd} made combined measurements of cross section with all decay modes.
Both combinations measure the production cross section by combining decay modes and also cross section times branching fraction for different production in each relevant decay mode. 
The ATLAS combination includes a total of 36 STXS bins with a precision down to 10\% level.
Overall, good agreement between data and SM prediction is observed.

\section{Differential Cross Sections}

Differential cross sections provide a more model-independent measurement in the fiducial region. 
The reconstruction-level distributions are unfolded~\cite{Schmitt_2017} to correct for efficiency loss, acceptance, and migration effects. 
Figure~\ref{fig:hyyhbb} (right) shows the differential cross sections in \pth\ measured by ATLAS in the $H \rightarrow WW^* \rightarrow e\nu\mu\nu$ channel~\cite{ATLAS:2023hyd}. 
A combination of $H \rightarrow ZZ^* \rightarrow 4\ell\ (\ell = e, \mu)$ and $H \rightarrow \gamma\gamma$ was performed by ATLAS~\cite{ATLAS:2022qef} to provide differential measurements with a better sensitivity, and the compatibility of the result with the SM prediction has a \textit{p}-value of 98\%.

%

\section{Di-Higgs}
\begin{figure}
\begin{minipage}{ \linewidth}
\centerline{\includegraphics[width=1 \linewidth]{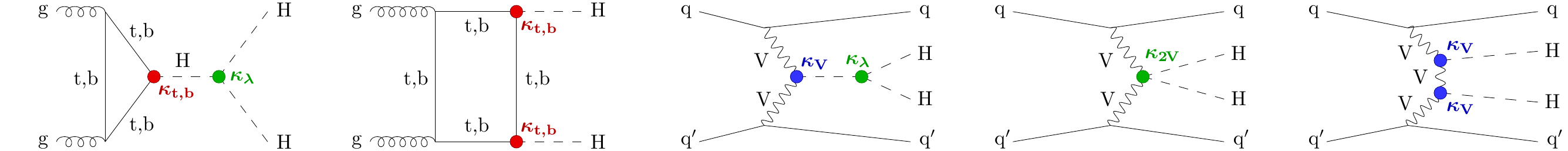}}
\end{minipage}
\caption[]{The main di-Higgs production modes ggF and VBF~\cite{CMS:2022dwd}.}
\label{fig:dihiggsProd}
\end{figure}

The di-Higgs processes (Figure~\ref{fig:dihiggsProd}) are sensitive to Higgs self-coupling \klambda\ through the ggF and VBF production modes, while also sensitive to two-Higgs-two-gauge-boson \ktwov\ through VBF mode. 

\begin{figure}
\begin{minipage}{0.24 \linewidth}
\centerline{\includegraphics[width=1 \linewidth]{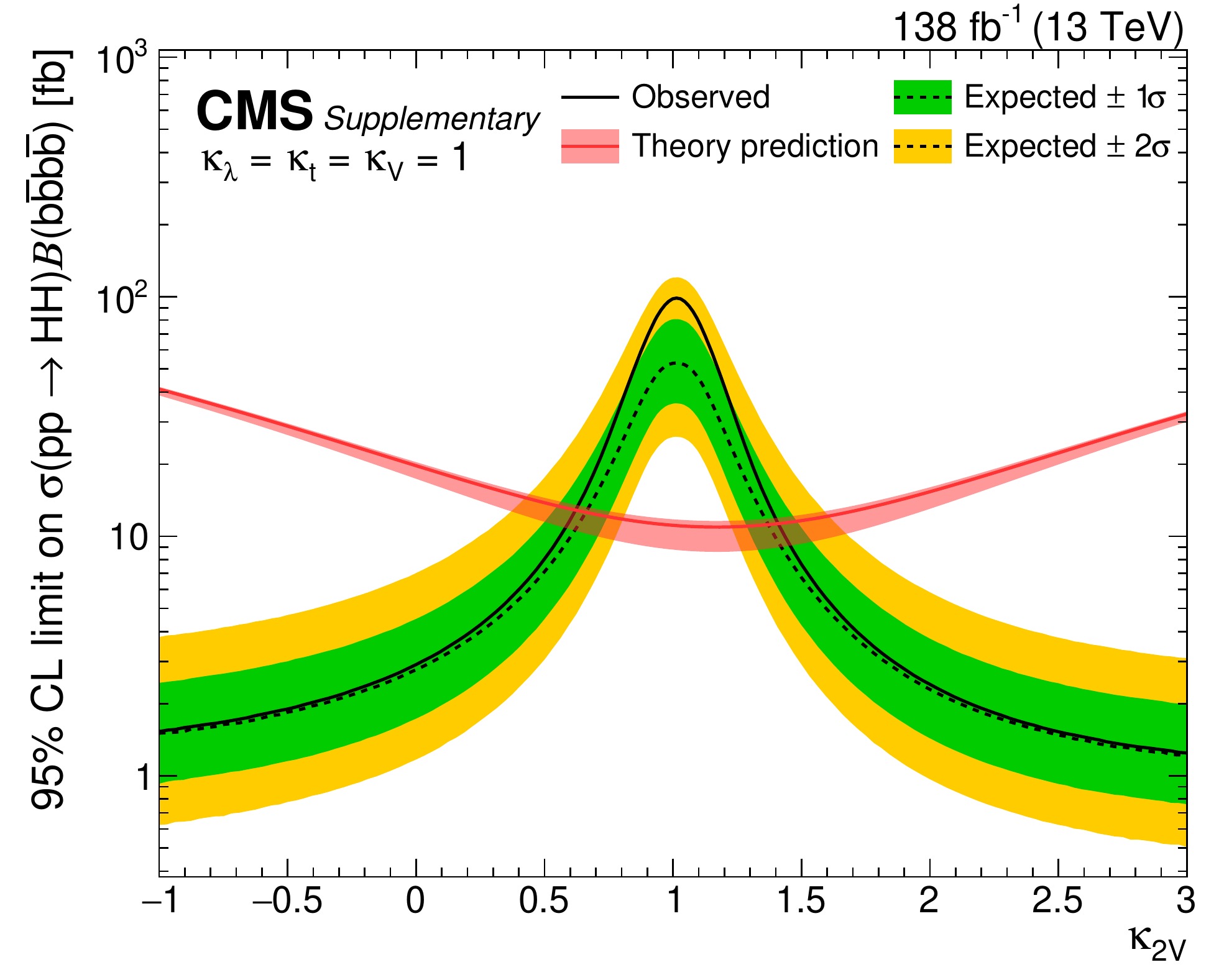}}
\end{minipage}
\hfill
\begin{minipage}{0.24 \linewidth}
\centerline{\includegraphics[width=1 \linewidth]{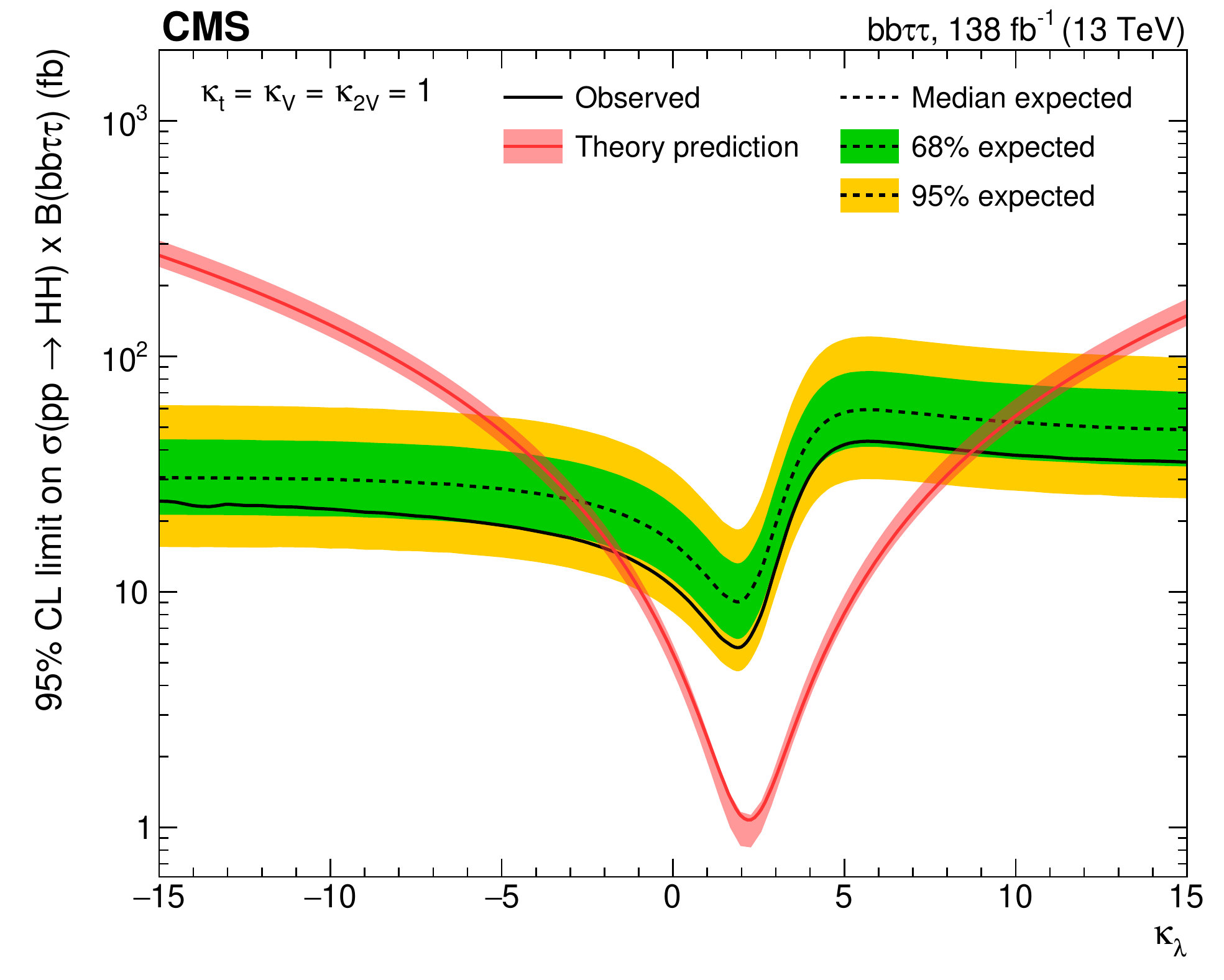}}
\end{minipage}
\hfill
\begin{minipage}{0.24 \linewidth}
\centerline{\includegraphics[width=1 \linewidth]{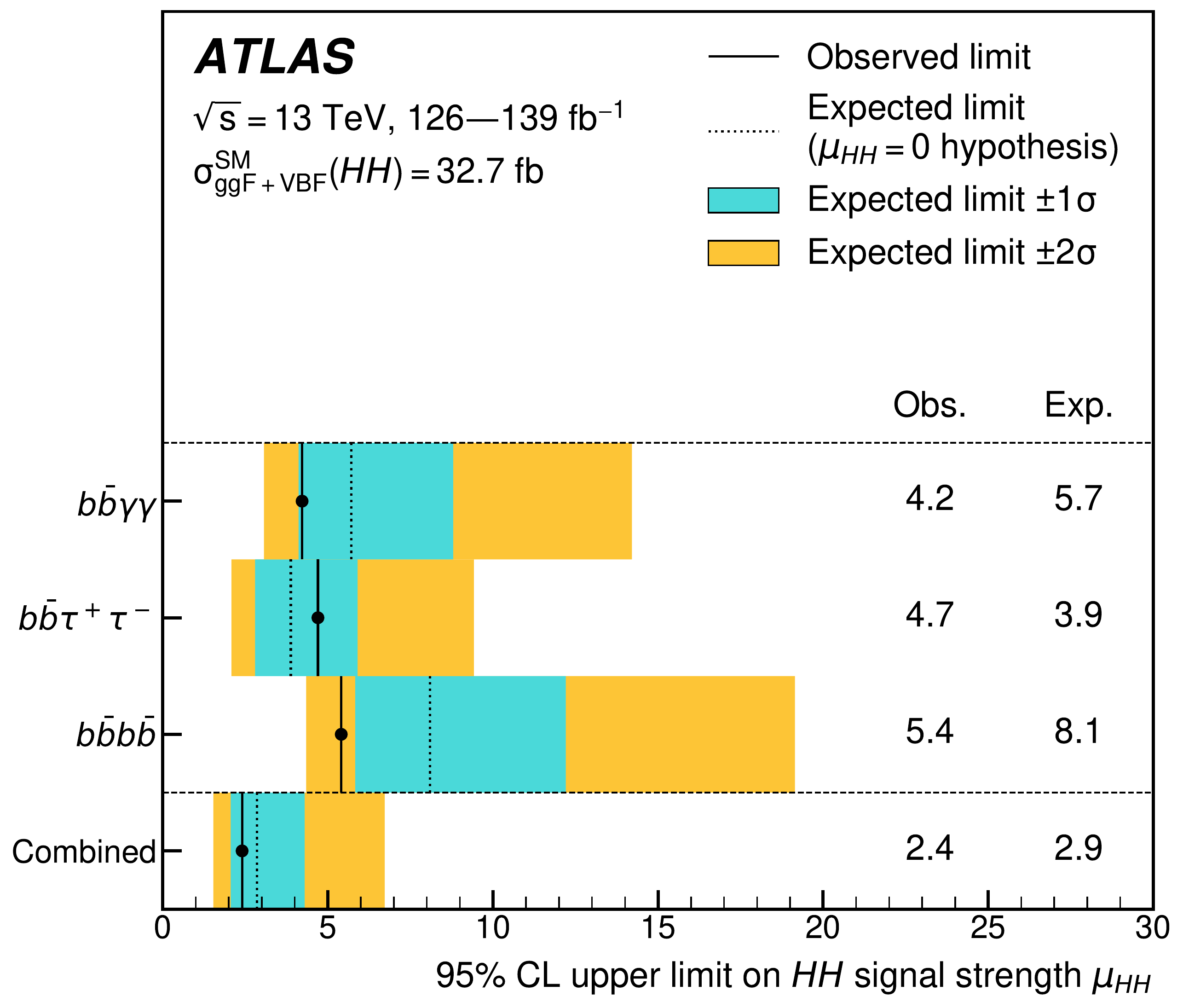}}
\end{minipage}
\hfill
\begin{minipage}{0.24 \linewidth}
\centerline{\includegraphics[width=1 \linewidth]{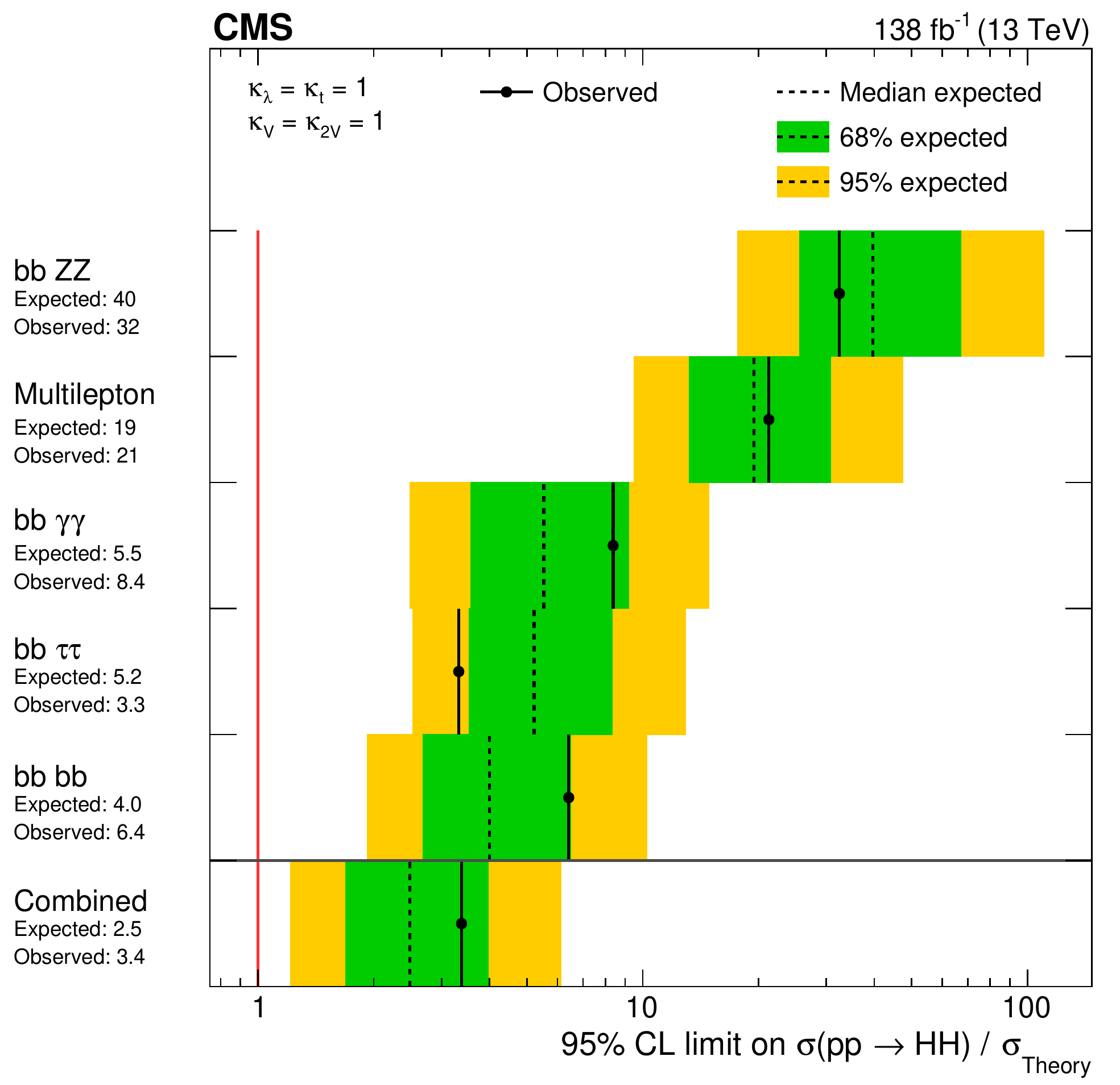}}
\end{minipage}
\hfill
\caption[]{95\% CL upper limits on production cross section as a function of \ktwov(left) in \textit{boosted} $HH \rightarrow 4b$ channel by CMS~\cite{CMS:2022nmn} and \klambda(middle left) in $H \rightarrow bb\tau\tau$ channel~\cite{CMS:2022hgz}. The 95\% CL upper limits on signal strength of HH production in combined analysis of ATLAS (middle right)~\cite{ATLAS:2022jtk} and CMS (right)~\cite{CMS:2022dwd}.   }
\label{fig:dihiggs}
\end{figure}

Di-Higgs processes are currently limited by data statistics.
The most sensitive channels are
4$b$, $bb\tau\tau$ and $bb\gamma\gamma$,
and include at least one $H \rightarrow b\bar{b}$ decay.
In the 4$b$ channel, the events could be categorized similarly to single Higgs studies into \textit{resolved} and \textit{boosted} topologies. 
In the \textit{resolved} topology, analyses are done by both ATLAS~\cite{ATLAS:2023qzf} and CMS~\cite{CMS:2022cpr}.
In the \textit{boosted} topology, CMS optimized a GNN algorithm, ParticleNet~\cite{PhysRevD.101.056019}, to select large-radius \textit{b}-jet and excluded $\ktwov = 0 $ with a significance of 6.3 $\sigma$ as in Figure~\ref{fig:dihiggs} (left)~\cite{CMS:2022nmn}.
The $bb\tau\tau$ analyses from both ATLAS~\cite{ATLAS:2022xzm} and CMS~\cite{CMS:2022hgz}  use multi-variate technique with data from all decay modes of $\tau\tau$.
The $bb\tau\tau$ is the most sensitive decay channel in all decay channels of di-Higgs for \textit{resolved} Higgs topology(Figure~\ref{fig:dihiggs} (middle left)). 
A combination of di-Higgs production is performed by both ATLAS~\cite{ATLAS:2022jtk} and CMS~\cite{CMS:2022dwd}. 95\% CL upper limits on the signal strength of di-Higgs production are set to be 2.9 and 3.4 times the SM prediction for ATLAS and CMS respectively.
Figure~\ref{fig:dihiggs} (middle right) and Figure~\ref{fig:dihiggs} (right) show the combined results for ATLAS and CMS respectively.

\section{Summary}

Higgs simplified template cross section measurements, differential cross section measurements and di-Higgs production cross section limits using the full Run 2 data collected by ATLAS and CMS experiments at the LHC were presented. 
The results are consistent with the Standard Model.
By the end of Run 3, the current dataset is expected to be doubled and will allow for a more granular measurement in kinematic regions for theory interpretation, and provide more stringent limits on HH production.

\section*{References}

\bibliography{WangRongkun} 

\end{document}